\documentclass[ floatfix, 12pt,nofootinbib, aps]{revtex4}
\usepackage{graphicx}
\usepackage{dcolumn}
\usepackage{bm}

\usepackage{amsmath,amssymb,amsthm}
\usepackage{braket}
\usepackage{hyperref}
\usepackage{color}

\begin{document}

\title{
The isotropic-to-nematic phase transition in hard helices: \\
theory and simulation
}

\author{Elisa Frezza, Alberta Ferrarini\footnote{alberta.ferrarini@unipd.it}}

\affiliation{
Dipartimento di Scienze Chimiche,
Universit\`{a} di Padova, via Marzolo 1, 35131 Padova, Italy
}

\author{Hima Bindu Kolli, Achille Giacometti\footnote{achille@unive.it}}

\affiliation{
Dipartimento di Scienze Molecolari e Nanosistemi,
Universit\`{a} Ca' Foscari di Venezia,
Dorsoduro 2137, 30123 Venezia, Italy
}

\author{Giorgio Cinacchi\footnote{giorgio.cinacchi@uam.es}}

\affiliation{
Departamento de F\'{i}sica Te\'{o}rica de la Materia Condensada,
Universidad Aut\'{o}noma de Madrid,
Campus de Cantoblanco, 28049 Madrid, Spain
}

\date{\today}

\begin{abstract}
\noindent
We investigate the isotropic-to-nematic phase transition in systems of hard helical particles, 
using Onsager theory and Monte Carlo computer simulations.
Motivation of this work resides in the ubiquity of the helical shape motif in many natural and synthetic polymers,
as well as in the well known importance that the details of size and shape have
in determining the phase behaviour and properties of (soft) condensed matter systems.
We discuss the differences with the corresponding spherocylinder phase diagram and 
find that the helix parameters affect the phase behaviour and the existence of the nematic phase.  
We find that for high helicity Onsager theory significantly departs from numerical simulations  
even when a modifed form of the Parsons-Lee rescaling is included  
to account for the non-convexity of particles.

\end{abstract}
\maketitle

\clearpage

\section{Introduction}

\noindent 
In the physics of fluids, simple and complex, it is well established that
size and shape of particles, either molecular or colloidal, play 
a key role in determining thermodynamics, structure and dynamics (e.g. Refs. \cite{BH,Barrat}).
Hard body particles, interacting with each other through  steep repulsive potentials only, can thus be viewed as elementary models to understand the
phase behaviour and properties of physical systems.

While the first example of phase transition driven by purely steric interactions is undoubtely the fluid-to-crystal phase 
transition in hard spheres \cite{earlyHS}, 
the isotropic-to-nematic liquid crystal (IN) phase transition in hard slender rods predicted by Onsager \cite{onsager} paved the way for an entirely new field.
Although Onsager theory was originally motivated by the observation of a nematic liquid crystal phase in suspensions of inorganic and biological rod-like colloidal particles \cite{oldexperiments}, its influence over the years have proven to be much more profund. 
The explanation of an ordered fluid formation, such as the nematic phase, as the result of
the competition between orientational and translational entropy contributions, introduced the 
far reaching concept of "ordering entropy", 
whilst the idea of expressing the system's free energy as a functional of single particle density is a precursor of modern density functional theory (DFT) (e.g. Ref. \cite{perspective}).

The original Onsager theory accounted only for the second-virial coefficient contribution, 
thus (strictly) limiting its  applicability to rod--like particle systems with large aspect ratios, 
but several improvements have been more recently proposed to overcome this drawback and 
include also higher order contributions. 
This prompted a number of DFT approaches with different degrees of sophistication,
as well as a series of computer simulations, \cite{acp93,lekkerkerker,tarazona2008}, 
that can be applied and extended to many systems, either mono- or poly-disperse, both homo- and hetero-geneous. 
All these studies have confirmed that entropy alone can lead to
complex phase organisations, including smectic and columnar liquid crystals.
Recent simulations have further unveiled  examples of the zoo of  morphologies that 
can be obtained from packing of particles of different shape \cite{glotzer,dijskra}. 

So far, most theoretical and computational studies have focussed on convex hard particles 
(e.g. Ref. \cite{acp93}), whilst 
concave particles have been given less attention.
Besides simple dumbbells, these include
bent-core \cite{allenbent,glaser},  lens-like \cite{lenti} and 
bowl-shaped \cite{bowls} particles. 
Somewhat surprisingly, hard helices are not part of the above list, in spite of the  
several examples of this shape that can be found in natural and synthetic polymers. 
 Rigid and semiflexible helical polymers (polynucleotides, polypeptides, viruses)  
have a well known propensity to form liquid crystal phases at high concentration \cite{PBLG,TMV,DNA,Hamley}. 
When examining and interpreting the experimental phase behaviour, 
helicoidal particles  were generally assimilated to rods,
thus neglecting  peculiarities related to the actual  shape (e.g. \cite{BelliniScience}).

\begin{figure}[htb]
   \centering%
   \includegraphics{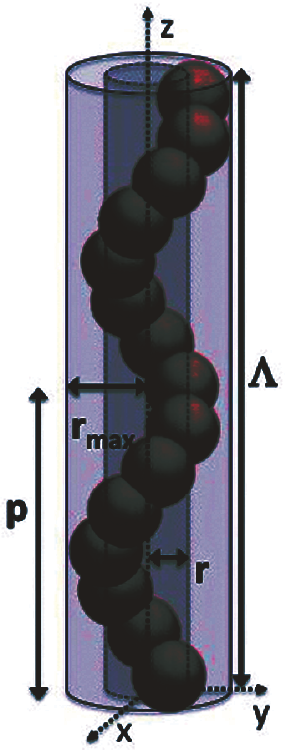}
   \caption {Model helix and its characteristic parameters.}
   \label{fig:helix}
\end{figure}

In this work, we address this problem by
undertaking a study of the phase behaviour of hard helices 
as a function of their structural features.
The model helices,  shown in  Fig. \ref{fig:helix}, are obtained by considering a set of fused hard spheres
all having diameter $D$, and arranged in a helical fashion along a string of contour length $L$. 
Helices of different shapes and Euclidean lengths $\Lambda$ are generated by tuning the radius $r$ and the pitch $p$
(the helix parameters are defined in Appendix A).

Considering systems of rigid homochiral helices, 
we have focussed on the effects of the particle shape on the IN phase transition.
Homochiral helices are expected to form a chiral nematic phase (cholesteric), 
in which the local preferred orientational axis (the director) rotates in space around a perpendicular axis.  However 
the cholesteric pitch is known to be orders of magnitude larger than length scale of inter-particle interactions; thus the local phase properties are virtually indistinguishable from those of the corresponding untwisted nematic phase and 
the influence of chirality on the phase boundaries can be neglected  at this stage \cite{chandrasekhar}.
In addition, the presence of periodic boundary conditions in the MC computations does not allow 
the emergence of an equilibrium cholesteric order, with a pitch much longer than the size of the simulation box.
For these reasons, phase chirality will not be considered in the present study and 
will be the object of a future dedicated study.

Our approach hinges upon an Onsager theory supported by Monte Carlo (MC) simulations.
We have used a simple extension of the Onsager theory suitable for particles with a finite aspect ratio, where the role of higher order virial terms is effectively taken into account 
\cite{parsons,lee}.
This theory was previously applied with success to the thermodynamics of the IN phase transition in hard, straight rods \cite{lee}.
Here its performance for the more subtle and challenging case of hard helices has been tested 
by comparison with MC simulations.
As further elaborated below, MC simulations for hard helices are considerably more demanding with respect to the spherocylinder counterpart.
Onsager theory, being computationally much less expensive, can then be rather useful for a preliminary exploration of the phase diagram.

Through a combined action of theory and simulations,
we will provide evidence that the helical shape significantly affects the location of the IN transition in comparison with
hard rods of similar length and diameter, thus casting some doubts on the possibility of providing a  straightforward link  
between the phase behaviour of hard helices and that of hard rods.

The layout of this work is as follows.
In the next section our Onsager and MC approaches are illustrated, while
theoretical and simulation results are presented and discussed in Section III.
Section IV concludes this work by summarising present findings and providing an outlook for future developments.

\section{Theory and simulations}

\subsection{Theory}

\subsubsection{Free energy of the isotropic and nematic phases}

\noindent
Let us consider a system of $N$ identical helices in a volume $V$ at temperature $T$.
We denote by $v=V/N$ the volume per particle and  $\eta=v_0/v$ the packing density, 
where $v_0$ is the volume of a particle.

The mutual interaction between a pair of hard helices (1 and 2) takes the form:
\begin{equation}
U(\bm{R}_{12},\bm{\Omega}_{12})=\left\{
\begin{aligned}
\infty & \quad \text{if 1,2 overlap} \\
0 & \quad  \text{if 1,2 do not overlap}
\end{aligned}
\right.
\label{eq:hard_inter}
\end{equation}
where $\bm{R}_{1}$ and $\bm{R}_{2}$ are the positions of the center-of-mass for helices $1$ and $2$ respectively, $\bm{R}_{12}=\bm{R}_{2}-\bm{R}_{1}$ is a vector defining the relative position of helix 2 with respect to 
helix 1 and $\bm{\Omega}_{12} = (\alpha_{12}, \beta_{12},\gamma_{12})$ are the Euler angles that 
define  the rotation from 1 to 2.

In the Onsager approach the free energy of the system is expressed as a functional of the single particle density function $\rho(\bm{R,\Omega})$, 
where $\bm{R}$ is the particle position and $\bm{\Omega}=(\alpha,\beta,\gamma)$ are 
the Euler angles specifying the particle orientation, 
with  the normalization condition
$\int d \bm{R} d\bm{\Omega} \rho(\bm{R,\Omega})=N$.
The nematic phase is uniform and the density function depends only on the particle orientation. 
Assuming helices as uniaxal particles,
the single particle density in the uniaxial nematic phase (see below) reduces to \cite{note1}: 
$\rho=\rho(\beta)=f(\beta)/ (4 \pi^2 v)$, 
where $\beta$ is the angle between the helix axis and the nematic director and $f(\beta)$ 
the orientational distribution function. 
The latter obeys the normalization condition  $\int_{-1}^{1} d(\cos\beta) f(\beta)=1$ and 
is a constant equal to 1/2 in the isotropic phase.

The free energy  can then be  expressed as:
\begin{equation}
A[f(\beta)]= Nk_BT\left[\ln \frac{\Lambda_{\text{tr}}^3}{V}
\frac{\Theta_{\text{or}}}{T} + \ln N -1 \right]
+A^\text{or}[f(\beta)]
+A^\text{ex}[f(\beta)]
\label{eq:A}
\end{equation}
The first term is the Helmholtz free energy of the ideal gas:
$\Lambda_{\text{tr}}=\left(h^2 / 2 \pi k_B T m \right)^{1/2}$ is the de Broglie wavelength and 
$\Theta_{\text{or}}=h^2/8\pi^2k_B I$ is the rotational temperature, with $k_B$ and $h$ being the Boltzmann and the Planck constant, respectively, 
while $m$ is the mass and $I$ is the inertia moment of the particle.
The second term in eq. \eqref{eq:A} accounts for the decrease 
of orientational entropy due to orientational ordering: 
\begin{equation}
\frac{A^\text{or}}{Nk_BT}=
 \int_{-1}^1  d(\cos\beta) f(\beta) \ln [2f(\beta)]
\label{eq:Aor}
\end{equation}
and the last term, $A^\text{ex}$, represents the excess free energy.
Within the Onsager formulation coupled with the Parsons-Lee (PL) correction \cite{parsons,lee},
this is expressed in terms of the second virial contribution with a pre-factor, $G(\eta)$, 
that is meant to account for higher virial contributions:
\begin{equation}
\frac{A^\text{ex}}{Nk_BT}= \frac{G(\eta)}{2 (4\pi^2)^2 v}
 \int d\bm{\Omega}_1\,f(\beta_1)  \int d\bm{\Omega}_2 \,f(\beta_2)
v_{\text{excl}}(\bm{\Omega}_{12})
\label{eq:Aexc0}
\end{equation}
where $v_{\text{excl}}(\bm{\Omega}_{12})$ is the volume excluded to a helix (2)  by another (1):
\begin{equation}
 v_{\text{excl}}(\bm{\Omega}_{12} )=- \int d \bm{R}_{12}\, e_{12}(\bm{R}_{12},\bm{\Omega}_{12})
\label{eq:vexc},
\end{equation}
with the Mayer function \cite{Barrat}:
\begin{equation}
 e_{12}(\bm{R}_{12},\bm{\Omega}_{12} )= \exp\{-U(\bm{R}_{12},\bm{\Omega}_{12} )/k_B T \}-1
\label{eq:Mayer}.
\end{equation}
Introducing the second virial coefficient:
\begin{equation}
 B_2={1 \over 2}\frac{1}{(4\pi^2)^2}
\int d\bm{\Omega}_1\, f(\beta_1) \int d\bm{\Omega}_2\,  f(\beta_2)
v_{\text{excl}}(\bm{\Omega}_{12})
\label{eq:B2}.
\end{equation}
the excess free energy eq. \eqref{eq:Aexc0}  becomes
\begin{equation}
A^{\text{ex}}[f]/N k_B T=G(\eta) B_2[f] /v.
\label{eq:AB2}
\end{equation}

\subsubsection{Parsons-Lee (PL) and Modified Parsons-Lee (MPL) approximations}
\noindent
The approximation proposed by Parsons \cite{parsons} and 
subsequently used by Lee \cite{lee}  and others \cite{mcgro,camp} for hard sperocylinders (and ellipsoids)  
relies on the assumption that the excess free energy is 
proportional to that of a system of hard spheres (HS) at the same packing fraction ($\eta$):
\begin{equation}
\frac{A^{\text{ex}}(\eta)}{Nk_BTB_2(\eta)}=\frac{A^{\text{ex}}_{HS}(\eta)}{Nk_BT B_2^{HS}(\eta)}
\label{eq:aex_PL0}
\end{equation}
Use of the Carnahan-Starling expression for the free energy of hard spheres \cite{Starling1969}, 
along with the relationships  $B_2^{HS}=4 v_{HS}$ and $\eta=v_{HS}/v$, 
where $v_{HS}$ is the volume of a hard sphere, yields
\begin{equation}
G(\eta)=\frac{A^{\text{ex}}_{HS}(\eta)}{Nk_BT B_2^{HS}(\eta)}=\frac{1}{4}\frac{4-3\eta}{(1-\eta)^2}
\label{eq:aex_PL1}
\end{equation}
In the original and subsequent works \cite{lee,mcgro,camp} 
the volume of the reference hard spheres was taken equal to that of the spherocylinders (or ellipsoids),  
$v_{HS}=v_0$. 
Good agreement between theory and simulations was obtained  in that case, but 
significant discrepancies were found  for linear particles made of tangentially bonded hard spheres \cite{jackson}.
It has then been suggested that the assumption $v_{HS}=v_0$ may be inappropriate for hard non-convex bodies, since in this case the free volume available at a given number density is smaller than for
convex particles having the same geometrical volume \cite{Varga2000}.
 It was proposed that in this case the volume of the reference hard spheres should be  replaced by an effective volume, $v_{\text{ef}}$, 
defined as the volume of the non-convex particle that is inaccessible to other particles. 
This effective volume is larger than the geometrical volume,  and for linear hard spheres
it has been evaluated in ref. \cite{Abascal1985} (see Appendix \ref{molvol}).
This variant of the PL theory has been given the name of modified Parsons-Lee (MPL) theory \cite{Varga2000}.

\subsubsection{Expansion in terms of orientational order parameters}
\noindent The orientational distribution function $f(\beta)$  is conveniently expanded in a basis of Legendre polynomials
\begin{equation}
f(\beta)=\frac{1}{2}\sum_{j=0}^\infty \,(4j+1)\braket{P_{2j}} P_{2j}(\cos\beta)
\label{eq:density_expansion}
\end{equation}
where
$\braket{P_{2j}}$ are the nematic order parameters 
\begin{equation}
\braket{P_{2j}} = \int_{-1}^1 d(\cos\beta) f(\beta) P_{2j}(\cos\beta)
\label{eq:order_parameter}
\end{equation}
The expansion is limited to polynomials of even rank in view of the nonpolar character of the nematic phase.
The order parameters take values in the range $-1/2 \leq \braket{P_{2j}} \leq 1$  and vanish in the isotropic phase.

 Upon substituting Eq.\eqref{eq:density_expansion} in Eqs. \eqref{eq:Aor}-\eqref{eq:Aexc0}
and exploiting the properties of Wigner rotation matrices \cite{Varshalovich},
we can express the orientational and excess contributions to the Helmholtz free energy as a function of the order parameters:
\begin{equation}
\frac{A^{\text{or}}}{Nk_B T}= \frac{1}{2}  \sum_{j=0}^\infty   \, (4j+1)   \braket{P_{2j}}
 \! \int_{-1}^1 \! d(\cos\beta) \! P_{2j}(\cos\beta) \! \  \ln \! \! \left [ \sum_{j^{\prime}=0}^\infty   (4j^\prime+1) \!  \braket{P_{2j^\prime}} \!  P_{2j^\prime}(\cos\beta) \! \right ]
\label{eq:aor_expansion}
\end{equation}
\begin{equation}
\frac{A^{\text{ex}}}{Nk_B T}=\frac{G(\eta)}{16 \pi^2v}  \sum_{j=0}^\infty (4j+1) 
\braket{P_{2j}}^2
\int  d\bm{\Omega}_{12} \, P_{2j}(\cos\beta_{12})  v_{\text{excl}}(\bm{\Omega}_{12})
\label{eq:aex_expansion}
\end{equation}
This leads to the following expression for the pressure
\begin{equation}
\begin{split}
\frac{P}{k_B T}=- \frac{1}{k_B T} 
\left(\frac{\partial A}{\partial V} \right)_{NT}=&\frac{1}{v}+\frac{1}{16\pi^2 v^2} \left( G(\eta)+\frac{\eta (5-3\eta)}{4(1-\eta)^3} \right) \\
& \times \sum_{j=0}^\infty (4j+1) \!  \braket{P_{2j}}^2
\int  d\bm{\Omega}_{12} \, P_{2j}(\cos\beta_{12})  v_{\text{excl}}(\bm{\Omega}_{12})
\end{split}
\label{eq:pres_expansion}
\end{equation}

\subsection{Monte Carlo simulations}
In order to test the theoretical predictions, 
we implemented Isothermal-Isobaric (NPT) MC simulations \cite{wood,king}  on a system of $N$ hard helices,
contained in cubic or orthorhombic computational boxes, with the usual periodic boundary conditions. 
Simulations were organised in cycles, 
 each consisting of 2$N$ attempted particle moves (a random translation and rotation) and a volume move.
Rotation trial moves were implemented either using the Barker-Watts \cite{Baker69} or  
the quaternions methods \cite{Allen87,Frenkel02}, 
finding a good consistency between them.
Volume moves were either performed scaling up or down the box in those
cases where cubic boxes were used or attempting to change a randomly selected
edge of the box in the other cases.
Being concerned with the IN phase transition only, 
we neglected from the outset all possible nuances necessary to properly account for other phases, 
such as twisted--nematic boundary conditions and variable--shape computational boxes.

\begin{figure}[htb]
   \centering%
   \includegraphics{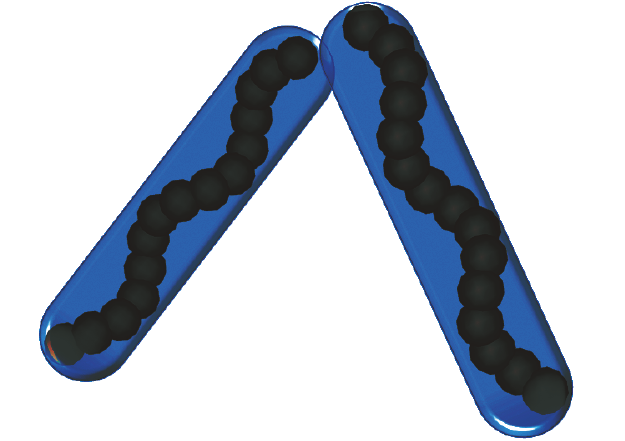}
   \caption {Cartoon of the overlap between the spherocylinders containg a pair of helices.}
   \label{fig:overlap}
\end{figure}

The overlap condition was computed by first inserting each helix into the smallest spherocylinder containing it and
testing for overlap between two such spherocylinders (see Fig.\ref{fig:overlap}).
This is a relatively fast test as it amounts to finding the minimal distance between two segments.

To this purpose, we used the algorithm proposed by Vega and Lago \cite{Vega94}. This method is approximatively four times
faster than others previously used, essentially because it reduces to only four the number of regions to be checked for
closest approach. 

Only in the event of overlap between two spherocylinders, 
the spheres forming the embedded helices were tested for overlapping. 
This procedure significantly reduced the computational cost  of the overlap test, 
that is one of the bottlenecks of this type of simulations, 
and considerably increased their efficiency.

The IN phase transition was monitored using the 
main orientational order parameter, $\braket{P_2}$, already discussed 
in the theoretical section. 
To this aim, the following tensor \cite{Veillard-Baron74}
\begin{eqnarray}
\label{mc:eq1}
\mathbf{Q}_{\alpha \beta} &=& \frac{1}{N} \sum_{i=1}^N \frac{3}{2} \mathbf{\widehat{u}}_{i}^{\alpha} \mathbf{\widehat{u}}_{i}^{\beta} -\frac{1}{2} \delta_{\alpha \beta}
\end{eqnarray}
was evaluated, and the corresponding eigenvalues and eigenvectors computed. 
Here $\alpha,\beta=x,y,z$ and $\mathbf{\widehat{u}}_{i}^{\alpha}$ is 
the $\alpha$ component of the unit vector $\mathbf{\widehat{u}}_{i}$ 
describing the orientation of the $i-$th helix axis. 
The orientational order parameter $\braket{P_2}$ was then identified with the largest eigenvalue of $\mathbf{Q}$. The difference between the other eigenvalues of $\mathbf{Q}$ was found to be smaller than 5$\%$, in agreement with our assuption of uniaxial nematic order.

\subsection{Computational details}
\subsubsection{Onsager theory calculations}
\noindent
For each system, the Helmholtz free energy  given in eqs. \eqref{eq:A}, \eqref{eq:aor_expansion} and \eqref{eq:aex_expansion}, 
is minimised at increasing values of the density $1/v$, 
and the order parameters of the stable phase at each density value  obtained 
are then used to calculate the pressure according to eq. \eqref{eq:pres_expansion}.

\paragraph{Evaluation of pair integrals}
\noindent Integrals over all the relative positions and orientations of pairs of particles, appearing in eqs. \eqref{eq:aex_expansion} and \eqref{eq:pres_expansion}, are preliminarly evaluated and stored, to be used for the calculations at the various density values. These integrals have the general form:
\begin{equation}
\int_0^{2 \pi} d \alpha_{12}   \int_{-1}^{1} d (\cos\beta_{12})  P_{2j}(\beta_{12})
\int_0^{2 \pi} d \gamma_{12}
\int_0^{2 \pi} d \phi_{12}   \int_{-1}^{1} d(\cos \vartheta_{12})  (R^0_{12})^3
\label{eq:int}
\end{equation}
where ${\bf{R}}_{12}$, the vector position of molecule 2 with respect to molecule 1, 
is expressed in spherical coordinates, ${\bf{R}}_{12} \equiv \{R_{12}, \phi_{12} ,\vartheta_{12} \}$ and $ R^0_{12}$ is 
the closest approach distance, which is a function of the relative position and orientation of the two molecules. 
The computational cost of this sixfold integral scales with $M^2$, where  $M$ is the number of spheres in a helix.
Gauss-Legendre and Gauss-Chebyshev quadrature algorithms are used  to evaluate these integrals  \cite{Numerical_Recipes}.

\paragraph{Free energy minimization}
\noindent
It is expedient to choose as variational parameters  the coefficients  $u_j$ of the expansion
\begin{equation}
-\ln f(\beta)= \sum_{j=1}^\infty u_j P_{2j}(\cos\beta),
\label{eq:u_mf_exp}
\end{equation}
rather than the order parameters. 
Eq. \eqref{eq:u_mf_exp} is used in eq. \eqref{eq:Aor} for the orientational contribution to the free energy, $A^{or}$, and is introduced into eq. \eqref{eq:aex_expansion} for the excess contribution, $A^{ex}$, through the  the order parameters,  eq. \eqref{eq:order_parameter}. Thus, the Helmholtz free energy is expressed as a function of the $u_j$ coefficients. This has a twofold advantage: The expansion eq. \eqref{eq:u_mf_exp} converges faster than that of the density function eq. \eqref{eq:density_expansion} and  
the parameters $ u_j$ are unconstrained, unlike order parameters.

\subsubsection{Monte Carlo simulations}

\noindent Our NPT MC simulations were carried out using $N=675$ or $867$ hard helices
with periodic boundary conditions.
As a general rule, we started a series of simulations from a diluted configuration  and
reached equilibrium upon compression. 
Typical equilibration runs consisted of $3 \times 10^6$ MC cycles and 
were followed by a production run of additional $3 \times 10^6$ MC cycles, 
during which averages of density and order parameter were calculated. 

In most of the equilibration runs the maximum values for
the displacement, rotation and volume moves were varied in the course
of the run to reach a 30-40\% of acceptance. 
This procedure is known to lead to a possible violation of the detailed balance condition
\cite{Miller00},
but we explicitly verified that this does not lead to any bias in the present case. 
During production runs, 
the overall
acceptance ratio was adjusted to be $30\%-40\%$ 
by a suitable choice of the maximum displacement, rotation and volume parameters,
and these values were never altered during the run.

It is worth emphasizing that simulations for hard helices are considerably more demanding 
from the computational point of view than simulations of  
hard spherocylinders. 
Depending on the state point considered and the values used for the radius and pitch, 
the computational cost might be as high as 8 times that of the corresponding spherocylinders.

\section{Results and Discussion}
\begin{figure}[htb]
   \centering%
   \includegraphics{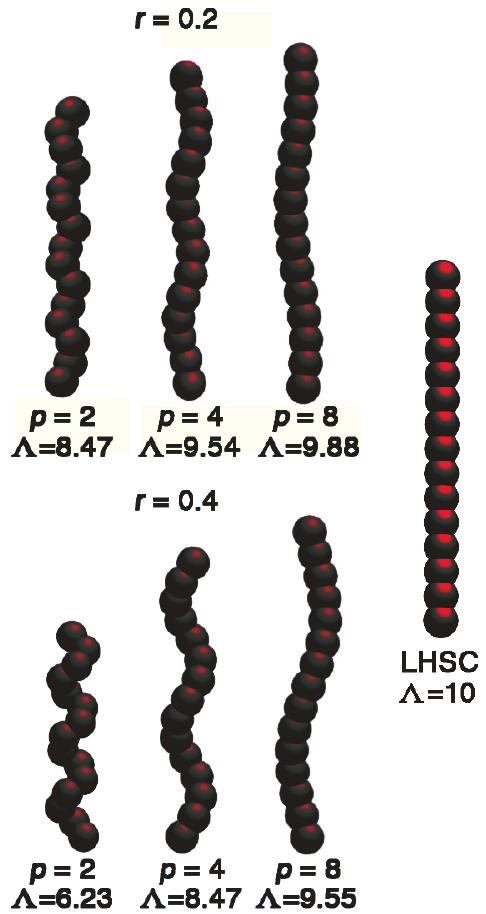}
\caption{Helices of radius $r$ and  pitch $p$ investigated here. Helices have the same contour length $L$ but different Eucliedean lengths $\Lambda$ (lengths are scaled with the the sphere diameter $D$). For comparison also the fully extended linear hard sphere chain (LHSC) is shown. }
   \label{fig:differenthelices}
\end{figure}

\noindent 
We have considered  helical particles with different structural parameters (see Appendix \ref{helix_model} for the definition),  as in Fig.\ref{fig:differenthelices}. Helices are formed by 15 fused spheres and have the same contour length $L=10D$, but with different pitches and radii.

In presenting and discussing our results we will use reduced units, with the diameter $D$ taken as the unit of length, and with reduced pressure $P^*={P D^3}/{k_B T}$. For each system, the results from MC simulations  were compared with those from Onsager theory with the PL and the MPL approximation, which differ in the definition of the packing fraction entering the scaling factor $G$, eq. \eqref{eq:aex_PL1}:  $\eta=v_0/v$, with $v_0$ being the geometric volume of the helix (PL), and $\eta=v_{ef}/v$,  where $v_{ef}$ is the effective volume defined in Appendix \ref {molvol} (MPL). Values of geometric and effective volume are reported in Tables \ref{tab:volumes1} and  \ref{tab:volumes}, respectively. 
MC data will be reported with error bars, evaluated according to the reblocking algorithm described in ref. \cite{Flyvbjerg}.
 
\begin{figure} [h!]
   \centering
   \includegraphics{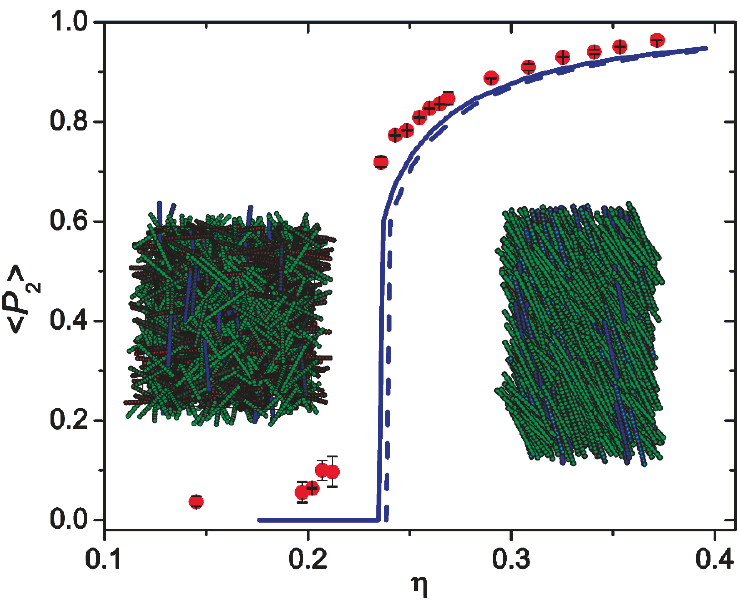}
   \qquad
    \includegraphics{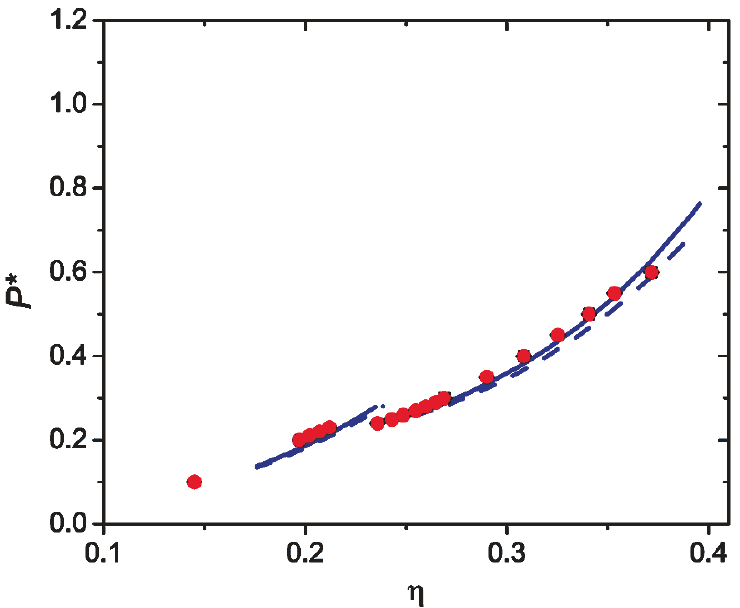}
   \caption { $\langle P_2 \rangle$  order parameter (Left) and reduced pressure $P^*$ (Right)  as a function of the volume fraction $\eta=v_0/v$ for the LHSC, from MC simulations (closed circles) and from Onsager theory with PL (dashed line) or MPL (solid line) approximation.  Insets on the left
panel, here and in following figures, depict representative snapshots obtained using QMGA software. \cite{QMGA}}
   \label{fig:p2_pres_pLHSC}
\end{figure}

As a preliminary test, we have performed calculations for the  
the linear hard sphere chain (LHSC), for which  $\Lambda$=$L$. 
Figure \ref{fig:p2_pres_pLHSC} shows order parameter $\langle P_2 \rangle$ and reduced pressure $P^*$ calculated for the LHSC as a function of the packing fraction  $\eta=v/v_0$.  
At $\eta \sim$ 0.24 an IN phase transition occurs, characterized by a jump in the order parameter. 
On moving deeper in to the N phase $\braket{P_2}$ takes higher values, larger than 0.8.
 The nonvanishing  $\langle P_2 \rangle$ obtained in the isotropic phase from simulations can be attributed to finite-size effects,
and this feature is also present in the isotropic phase for helices.
  Fig. \ref{fig:p2_pres_pLHSC} shows  good agreement between theory and simulations for  LHSCs. The results obtained using the PL and the MPL approximation are also very close one another, as expected in view of the high superimpositions of the spheres, so that the cavities between them have tiny volumes. This agrees with ref. \cite{Varga2000} where it was shown that for LHSCs the discrepancies between MC simulations and PL theory, and correspondingly also the improvements of the MPL scaling, decrease as the superposition between adjacent spheres increases.

\begin{figure} [h!]
   \centering
   \includegraphics{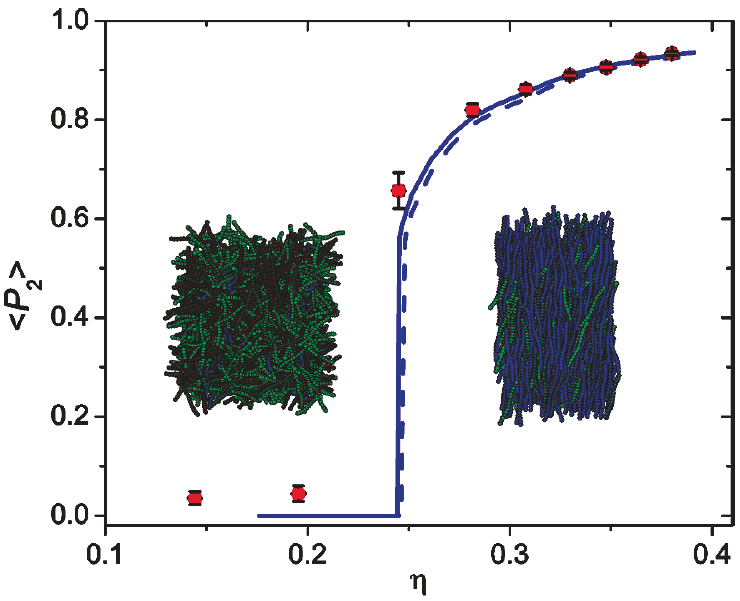}
   \qquad
    \includegraphics{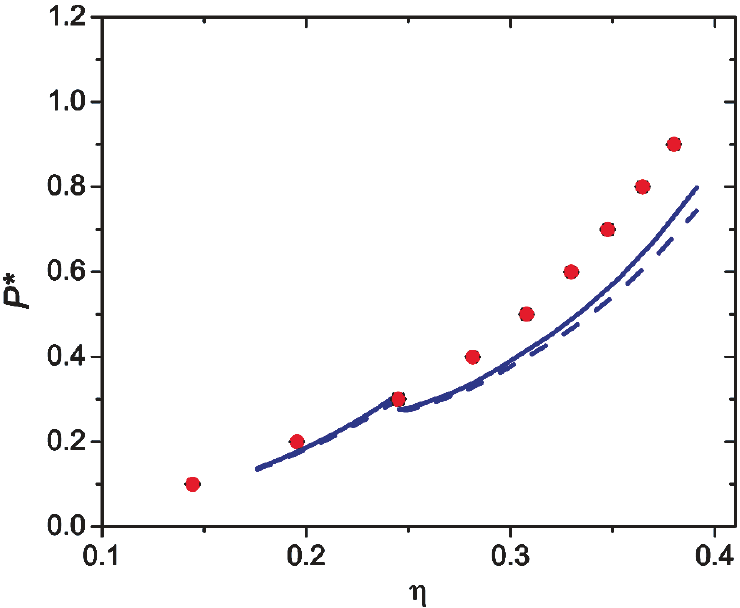}
   \caption {$\langle P_2 \rangle$ order parameter (Left) and reduced pressure $P^*$ (Right) as a function of the volume fraction $\eta=v_0/v$  for the  helix with $p$= 8 and $r$= 0.2,  from MC simulations (closed circles) and from Onsager theory with PL (dashed line) or MPL (solid line) approximation.}
   \label{fig:p2_pres_p8r02}
\end{figure}

\begin{figure} [h!]
   \centering
   \includegraphics{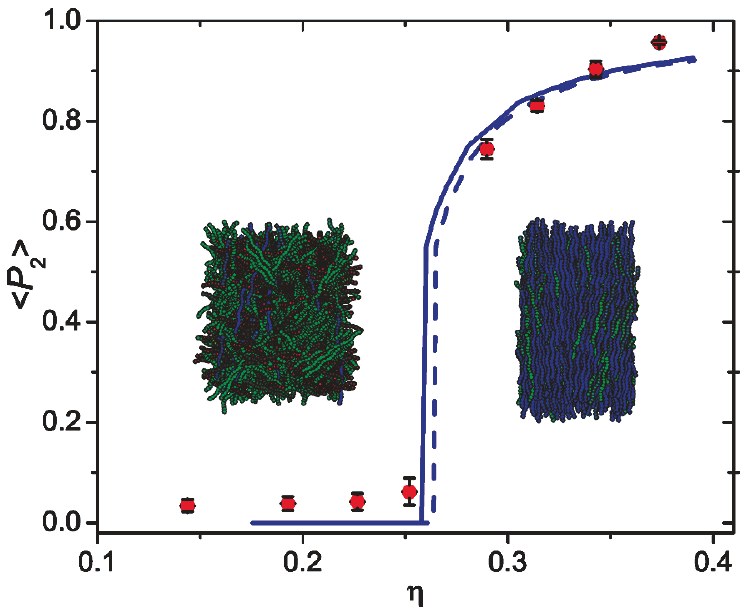}
   \qquad
    \includegraphics{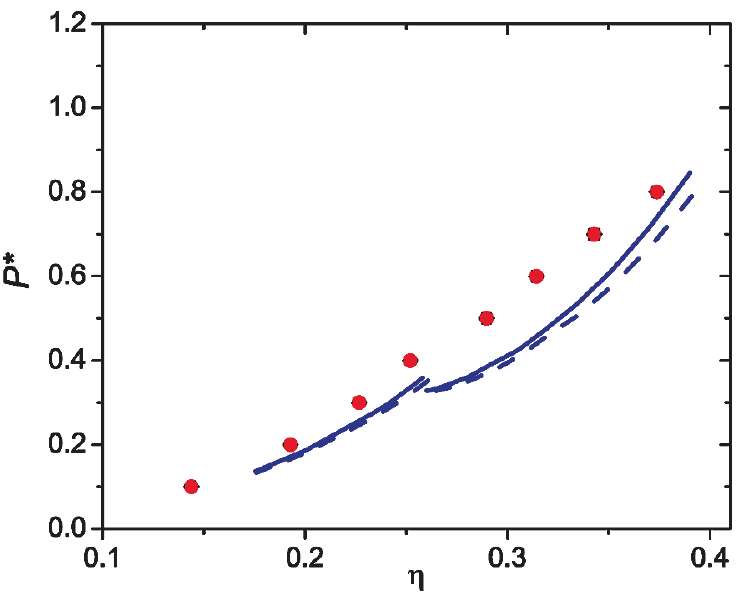}
   \caption {$\langle P_2 \rangle$ order parameter (Left) and reduced pressure $P^*$ (Right) as a function of the volume fraction $\eta=v_0/v$ for the  helix with $p$= 4 and $r$= 0.2, from MC simulations (closed circles) and from Onsager theory with PL (dashed line) or MPL (solid line) approximation.}
   \label{fig:p2_pres_p4r02}
\end{figure}

\begin{figure} [h!]
   \centering
   \includegraphics{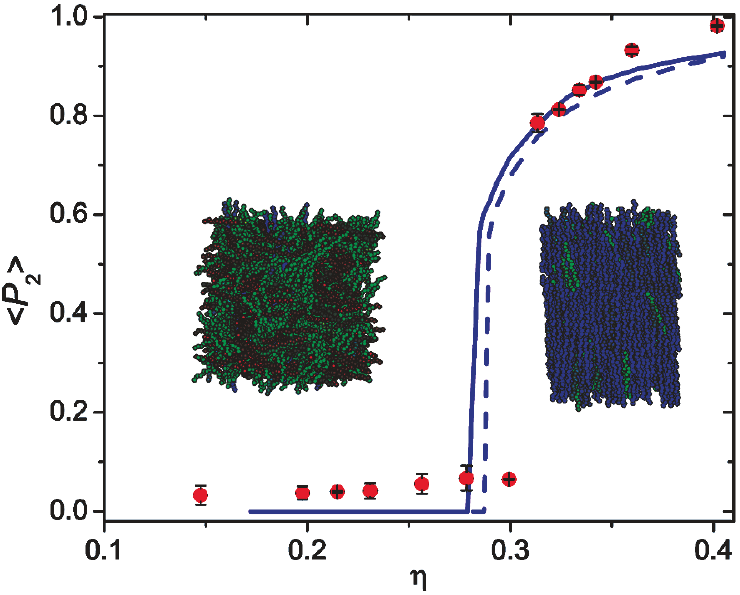}
   \qquad
    \includegraphics{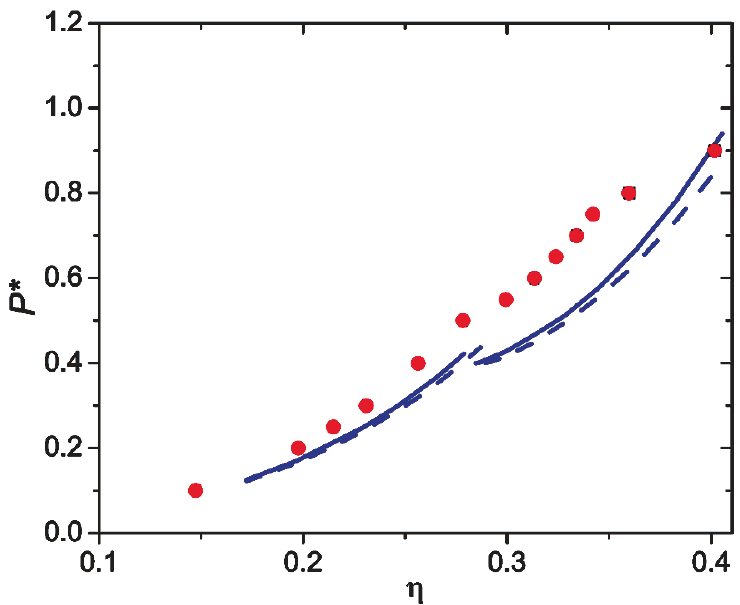}
   \caption {$\langle P_2 \rangle$ order parameter (Left) and reduced pressure $P^*$ (Right) as a function of the packing fraction $\eta=v_0/v$ for the  helix with $p$=2 and $r$= 0.2, from MC simulations (closed circles)  and from Onsager theory with PL (dashed line) or MPL (solid line) approximation.}
   \label{fig:p2_pres_p2r02}
\end{figure}

Figures \ref{fig:p2_pres_p8r02}-\ref{fig:p2_pres_p2r02}  show order parameters and pressures calculated for the  helices with $r$=0.2 and  decreasing pitches $p$=8, 4 and 2.
In all these cases a IN transition  is clearly visible, with its location in densities shifting  from $\eta\sim 0.24$ to $\eta\sim 0.29$ with decreasing pitches from LHSC (infinite pitch) to the helix with shorter pitch ($p=2$). This can be qualitatively understood in terms of 
the decrease of the Euclidean length (and hence the aspect ratio) with decreasing pitch. In all these cases, we find a good agreement between Onsager theory and numerical simulations in the location of the IN transition and in the density dependence of the $\braket{P_2}$ order parameter.
However pressure tends to be  understimated by theory, especially in the N phase, and  this differences increase with increasing density and with decreasing pitch. 
The PL approximation does not appear to be adequate for these helical particles and use of the MPL variant leads only to a very slight improvement. 
The reason is that the non-convexity of the helices is not simply due to the voids between adjacent spheres (see Fig. \ref{fig:spheres}), so
removal of these voids is not sufficient to account for the the real excluded volume.

The discrepancy with respect to LSHCs becomes even more pronunced for larger radii, as depicted in 
Figures \ref{fig:p2_pres_p8r04}-\ref{fig:p2_pres_p2r04}  reporting the $\langle P_2 \rangle$ order parameter and the reduced pressure calculated for helices with  $r$=0.4 and $p$=8, 4 and 2. These helices are curlier than those with smaller $r$ value (see Fig. \ref{fig:differenthelices}), so it is not surprising that the differences from the behaviour of LSHCs are even more pronounced. 
No clear N phase is observed in simulations for the helices with $p$=4 and $p$=2, although at sufficiently high packing fraction ($\eta \approx 0.35$) an anisotropic organization, with some  signature of layered ordering, is visible. A complete characterization of these phases is delicate, mainly due to equilibration problems, and  is presently under scrutiny. In the case of $p=8$ (see Fig.\ref{fig:p2_pres_p8r04}), a nematic phase was detected between $\eta \sim 0.27$ and $\eta \sim 0.38$; interestingly, the IN transition occurs at higher density than for the helices with smaller radius and similar Euclidean length ($r$=0.2 and $p$=4). 
 In all helices with $r$=0.4 we have also found  a marked deviation between theoretical and MC results. In contrast to simulations, a nematic phase is predicted by Onsager theory for all pitch values, with the IN transition occurring at increasing density as the pitch decreases. Of course, being the theory  implemented only for isotropic and uniaxial nematic phases, other possible phases could not be investigated.
 In short, only for the most elongated system ($r$=0.4 and $p$=8), we find a reasonable agreement between theory and simulations in this case. For shorter pitches, a jump 
in the {$\langle P_2 \rangle$ order parameter is obtained from simulations and theory at similar $\eta$ values,  but the ordered phases appear to be different.  As for pressure, differences between theory and simulations  even appear  in the isotropic phase for the helices with $p$=4 and $p$=2,  with theoretical predictions lower than the MC results.  For $p$=2 the improvement deriving from the MPL approximation is more significant than in the other cases, due to the larger value of the effective volume determined for this system using the rolling sphere criterion (see Table \ref{tab:volumes} and Appendix \ref{molvol}).
\begin{figure} [h!]
   \centering
   \includegraphics{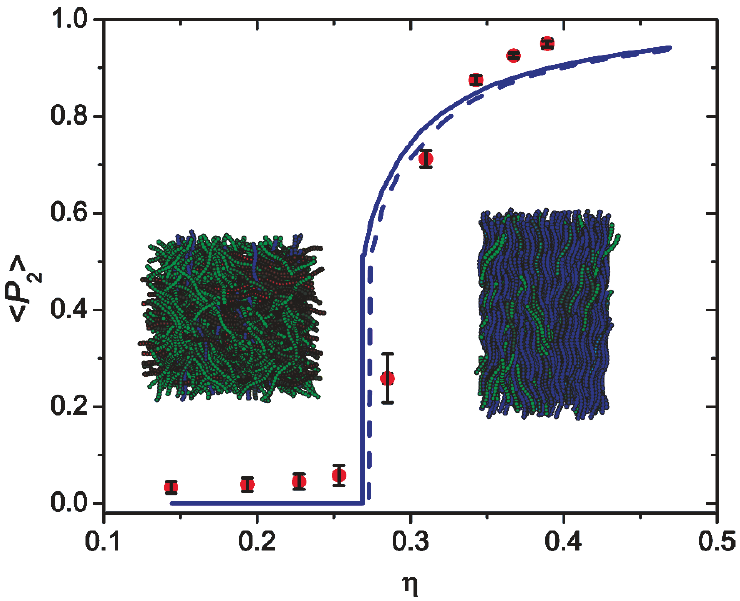}
   \qquad
    \includegraphics{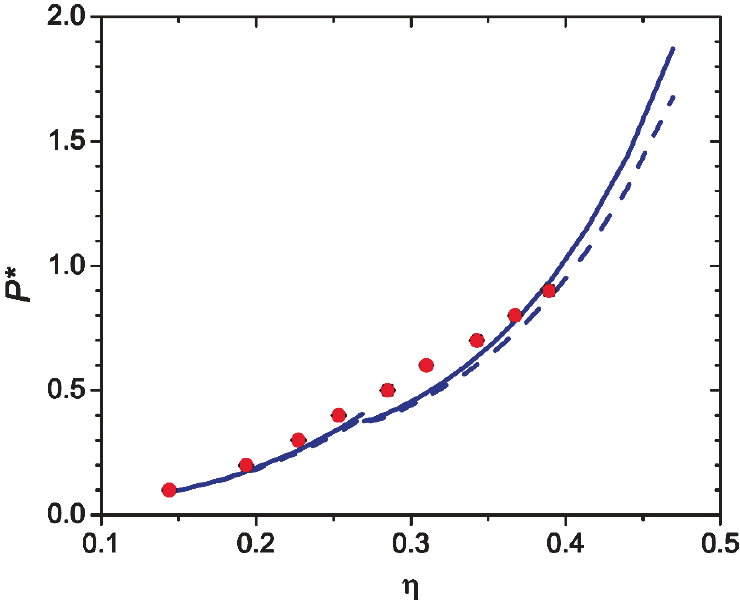}
   \caption {$\langle P_2 \rangle$ order parameter (Left) and reduced pressure $P^*$  (Right) as a function of the volume fraction $\eta=v_0/v$ for the  helix with $p$= 8 and $r$= 0.4, from MC simulations (closed circles) and from Onsager theory with PL (dashed line) or MPL (solid line) approximation.}
   \label{fig:p2_pres_p8r04}
\end{figure}
\begin{figure} [h!]
   \centering
   \includegraphics{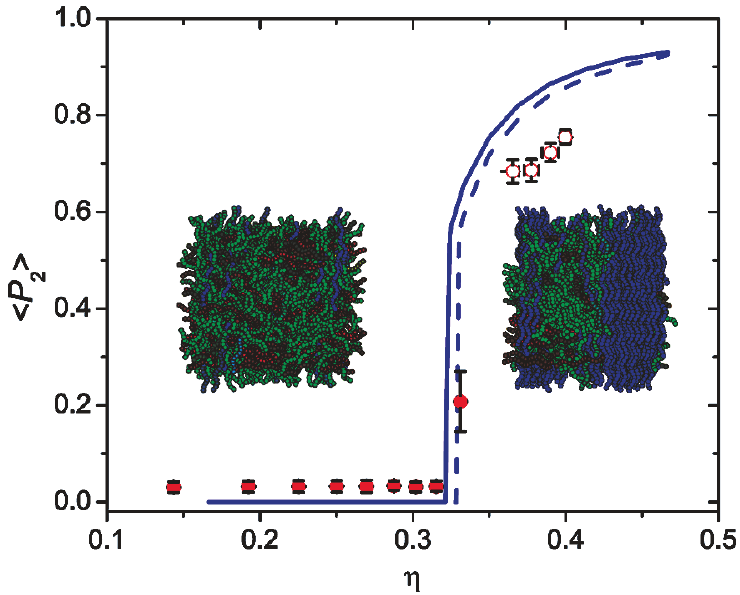}
   \qquad
    \includegraphics{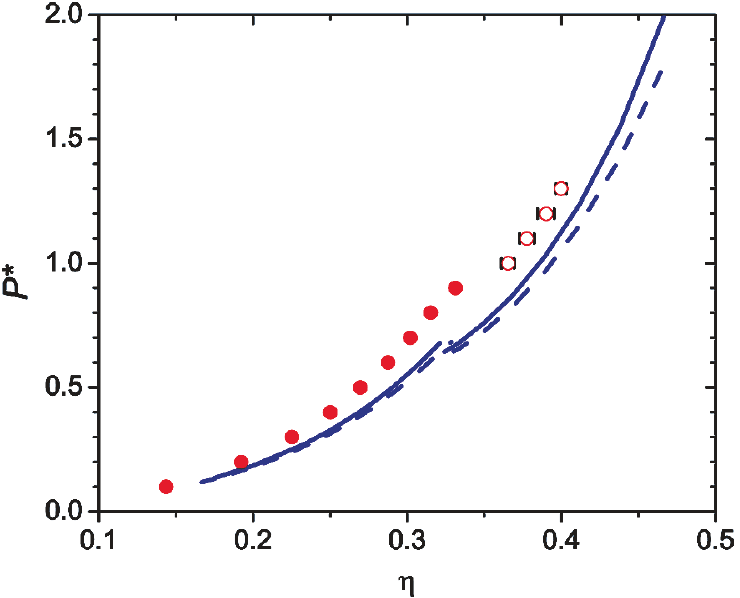}
   \caption {$\langle P_2 \rangle$ order parameter (Left) and reduced pressure $P^*$  (Right) as a function of the volume fraction $\eta=v_0/v$ for the  helix with $p$= 4 and $r$= 0.4, from MC simulations (circles) and from Onsager theory with PL (dashed line) or MPL (solid line) approximation. Open circles are used for metastable states not yet fully characterized.}
   \label{fig:p2_pres_p4r04}
\end{figure}

\begin{figure} [!h]
   \centering
   \includegraphics{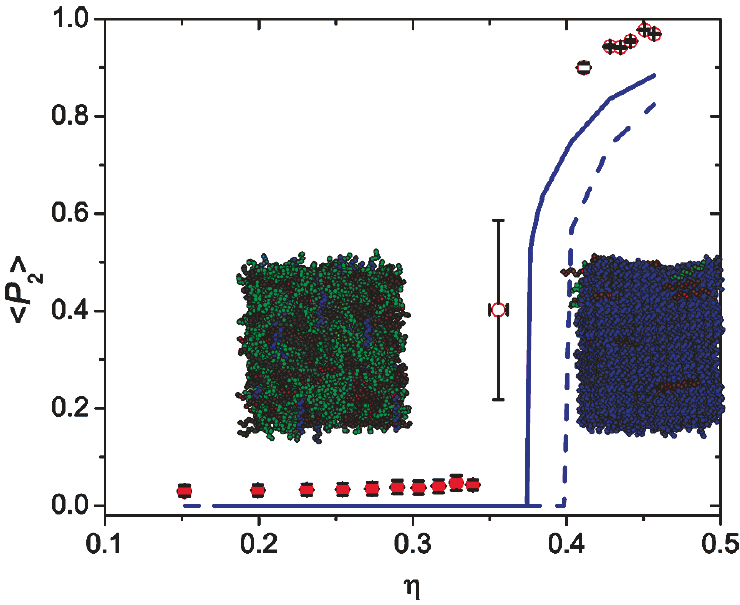}
   \qquad
    \includegraphics{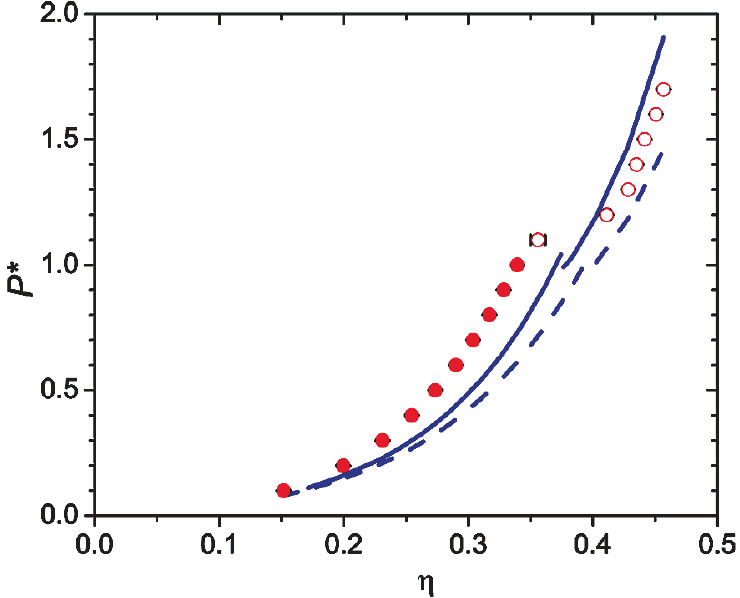}
   \caption {$\langle P_2 \rangle$ order parameter (Left) and reduced pressure $P^*$ (Right) as a function of the volume fraction $\eta=v_0/v$ for the  helix with $p$= 2 and $r$= 0.4, from MC simulations (circles) and from Onsager theory with PL (dashed line) or MPL (solid line) approximation. Open circles are used for metastable states  not yet fully characterized.}
   \label{fig:p2_pres_p2r04}
\end{figure}

 An interesting last point, related to the above findings, is whether 
the IN phase transition for helices can be mapped on to that of rods in terms of simple parameters like the aspect ratio, 
as generally done in experimental work on helical systems \cite{BelliniScience}.
Figure \ref{fig:all} collects the theoretical predictions of the IN phase transition as a function the Euclidean length $\Lambda$. 
For comparison, the results obtained for LHSCs and those for spherocylinders are also reported.
In the latter case the Onsager expression for the excluded volume was used \cite{onsager}. 
Of course the contour length $L$, which is identical for all the helices, is not 
a significant parameter in relation to the IN phase transition.  On the other hand, Figure \ref{fig:all} also suggests that 
the Euclidean length, although more meaningful, is not fully satisfactory either, since 
for the same aspect ratio $\Lambda/D$, 
 the density at which the IN transition occurs has a non trivial dependence on the combination of the helical parameters $r$ and $p$.
As a general rule, we find  the transition to move towards higher volume fraction with increasing degree of non-convexity. The fact that the location of the IN phase transition is not uniquely related to the aspect ratio may have implications for the analysis of experimental data for helical particles, as anticipated.

\begin{figure}[h!]
   \centering%
   \includegraphics{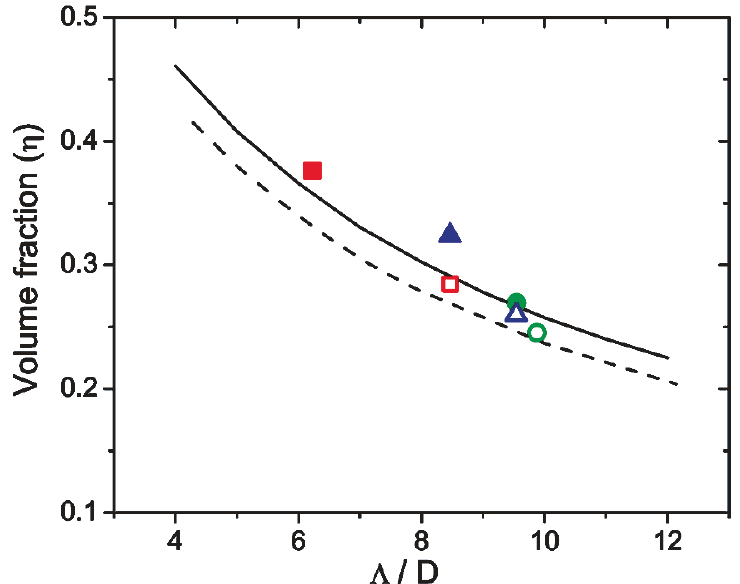}
   \caption {Volume fraction $\eta=v_0/v$ at the isotropic-nematic transition as a function of the Euclidean length ($\Lambda$), obtained from the Onsager theory with MPL approximation. Symbols refer to helices of different radius $r$ (open for $r= 0.2$ and closed for $r=0.4$) and pitch $p$ (squares for $p=2$, triangles for $p=4$ and circles for $p=8$ ), lines are for LHSCs (dashed) and spherocylinders (solid).}
   \label{fig:all}
\end{figure}

\section{Conclusions}
\noindent
In this work,  a system of hard helical particles have been investigated
using an  Onsager theory and MC computer simulations.
Our  main goal was to rationalize the changes in the isotropic-to-nematic phase transition 
on going from straight rod-like to quite tortuous helical particles. 
We have found that helicity affects the location of the IN phase transition, 
with the  latter in general  being shifted to higher densities with increasing aspect ratios, 
as in spherocylinders. 
However, the same aspect ratio can be achieved with different structural parameters of hard helices, 
and this affects the IN  phase transition.
In other words, the aspect ratio alone cannot be considered as a good candidate for the interpretation of liquid-crystal phase diagrams of strongly curled helical particles.
Our numerical results also unveiled the presence of additional ordered phases, 
especially in the case of highly distorted partices, that will require further analysis to assess their specific natures.
 
Another objective of our study was  a test of  Onsager theory for helical systems.
We have examined whether the Onsager theory,  
which has proved to be  successful in accounting for the thermodynamics of the IN phase transition in hard rods, 
can be extended to systems of helical non-convex particles. 
We have found that for high helicity Onsager theory significantly departs from numerical simulations, 
even when a modifed form of the Parsons-Lee rescaling is included to account for the non-convexity of particles.
When compared to the MC simulation data, Onsager theory generally understimates pressure, 
with deviations that increase with increasing density and upon going from the isotropic to the nematic phase.
This points to the need of a more effective theory for hard non-convex particles, 
a field that remains largely unexplored.
Besides the Onsager theory employed here, 
various other theoretical approaches have been proposed, which include 
scaled-particle theory \cite{Cotter,Boublik,Wu}, 
the Vega and Lago theory that aims at incorporating a better description of the isotropic state \cite{VegaJCP}, 
as well as Wertheim statistical mechanical treatment of associating fluids \cite{Wertheim}, 
which was successfully applied to bent-shaped particles in the isotropic phase \cite{camp}. 
Another approach envisages the extension of the Onsager theory beyond the second virial contribution. 
We intend to undertake a thorough analysis and comparison of available theories in a future work.

This study is preliminary from several viewpoints and we  plan to extend it in a number of ways, as alluded before. 
Though we dealt with chiral particles, phase chirality was not considered here, 
and the relationship between particle and phase chirality is one of the high priority points in our agenda. 
 Other points that deserve close attention, and are currently under investigation, are a detailed numerical definition of the IN coexistence and the characterization of other phases occurring at higher densities.

\acknowledgments
\noindent E.F. acknowledges the University of Padova for a scholarship (PhD School in Materials Science and Engineering). H.B.K. acknowledges MIUR for a PhD scholarship.  A.F. and A.G acknowledge financial support from PRIN-COFIN2010-2011 (contract 2010LKE4CC). G.C. is grateful to the Government of Spain for the award of a Ram\'{o}n y Cajal research fellowship.

\appendix

\section{Definition of fused hard sphere helices}
\label{helix_model}
\noindent A helix is made of $N$ spheres, whose centers are located at the points defined by the parametric equations:
\begin{equation}
\left\{
\begin{aligned}
x_i&=r \cos(2{\pi}t_i) \\
y_i&=r \sin(2{\pi}t_i) \qquad \qquad 1 \leq i \leq N \\
z_i&=pt_i
\end{aligned}
\right.
\label{eq:helix}
\end{equation}
where $r$ is the radius and $p$ is the pitch of the helix (see Fig. \ref{fig:helix}).
Given the values of $r$, $p$ and of the contour length $L$, the increment $\Delta t =t_{i+1}-t_i$ is determined by the equation:
\begin{equation}
\frac{L}{N-1}=2\pi\Delta t \sqrt{r^2+\left(\frac{p}{2\pi}\right)^2}.
\end{equation}
The Euclidean length of the helix is defined as $\Lambda=z_{N}-z_1$, depends on the pitch and radius, and coincides with  the contour length $L$ only for $r=0$.

\section{Molecular volume and effective volume of helices}
\label{molvol}
\noindent The volume $v_0$ of a linear chain formed by $m$ fused hard spheres (LHSC) of diameter $D$ and center-to-center distance  $d_{cc}$ (see Fig. \ref{fig:spheres}) is given by
\begin{equation}
v_{0}=\frac{\pi}{6}D^3\left[1+\frac{m-1}{2} \left(3\frac{d_{cc}}{D}- \left(\frac{d_{cc}}{D}\right)^3\right)\right]
\label{v0}
\end{equation}
The same expression holds for a helix of fused hard spheres, provided that there are only two-sphere overlaps and the correct value of the distance $d_{cc}$ is used.\footnote{$d_{cc}=\sqrt{(x_{i+1}-x_i)^2+(y_{i+1}-y_i)^2+(z_{i+1}-z_i)^2}$ is the distance between centres of two subsequent spheres.} For a given length of the curve connecting the centers of a pair of subsequent spheres, this distance depends on the helix radius and pitch. Table \ref{tab:volumes1} reports the (geometric) volume calculated for the all the helices shown in Fig. \ref{fig:differenthelices}. 

\begin{table}[h]
\centering
\begin{tabular}{ccc}
\toprule
helix & $ {d_{cc}}$ & $ {v_0} $  \\
\colrule
$p$= 2, $r$= 0.2 & 0.687 & 6.89  \\
$p$= 2, $r$= 0.4 & 0.680 & 6.85   \\
$p$= 4, $r$= 0.2 & 0.711 & 7.02  \\
$p$= 4, $r$= 0.4 & 0.707  & 7.00  \\
$p$= 8, $r$= 0.2 & 0.714 & 7.04  \\
$p$= 8, $r$= 0.4 & 0.714  & 7.04  \\
LHSC &  0.714  & 7.04  \\
\botrule
 \end{tabular}
\caption{Volume $v_0$ 
of fused hard sphere helices of radius $r$ and pitch $p$, calculated using Eq. \eqref{v0}. For comparison, also the value for LHSCs is reported.}
\label{tab:volumes1}
\end{table}

A definition of the \emph{effective} volume has been proposed for LHSCs, as the volume enclosed by the surface drawn by a sphere identical to those of the chain, rolling over the particle \cite{Abascal1985}. 
An example of this surface is shown in Fig.\ref{fig:spheres}. The effective volume of the LHSC is then  given by the expression:
\begin{equation}
v_{\text{ef}}^{\text{LHSC}}=\frac{\pi}{6}D^3\left[1+(m-1)\! \! \left(3\frac{d_{cc}}{D}-\frac{1}{2}\left(\frac{d_{cc}}{D}\right)^3 \!- 3\sqrt{ \left(1-\left(\frac{d_{cc}}{2D}\right)^2\right)}\arcsin\left(\frac{d_{cc}}{2D}\right)\right)\right]
\label{vef}
\end{equation}
\begin{figure} [htbp !]
  \centering
   \includegraphics{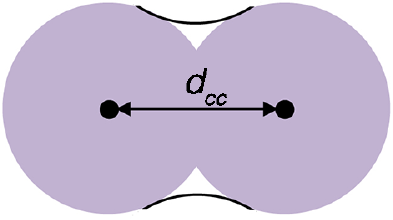}
  \caption {Surface defining the effective volume of a pair of fused hard spheres.}
   \label{fig:spheres}
\end{figure}

We have adopted the same definition of the effective volume for fused hard sphere helices. However in this case, depending on the helix curvature, the effect of the rolling sphere can go beyond that of simply filling the voids between subsequent beads. We have calculated the effective volume ($v_{\text{ef}}$) of helices using the program MSMS \cite{MSMS}. The rolling sphere radius was taken equal to the radius of the fused hard spheres that form the helix.
Table \ref{tab:volumes} reports the $v_{\text{ef}}$ values obtained for the helices shown in Fig. \ref{fig:differenthelices}; for comparison we report in the table also the volume calculated according to eq. \ref{vef}, using for each sphere  the appropriate $d_{cc}$ value ($v_{\text{ef}}^{\text{LHSC}}$). We can observe that   $v_{\text{ef}}=v_{\text{ef}}^{\text{LHSC}}$  for all helices with longer pitch; only for $p=2$ there is some difference,  more pronounced in the case with $r=0.4$. This discrepancy can be understood considering  that these helices have  grooves narrower than the sphere diameter $D$ (see Fig.\ref{fig:spheres}).

\begin{table}
\centering
\begin{tabular} {ccc}
\toprule
helix & $ v_{\text{ef}}$ & $v_{\text{ef}}^{\text{LHSC}}$\\
\colrule
$p$= 2, $r$= 0.2 & 7.24 &  7.20  \\
$p$= 2, $r$= 0.4 & 7.78 &  7.15  \\
$p$= 4, $r$= 0.2 & 7.37 &  7.37  \\
$p$= 4, $r$= 0.4 & 7.34 &  7.34  \\
$p$= 8, $r$= 0.2 & 7.39 &  7.39  \\
$p$= 8, $r$= 0.4 & 7.39 &  7.39  \\
LHSC &  7.39  & 7.39  \\
\botrule
\end{tabular}
\caption{Effective volume 
of fused hard sphere helices of radius $r$ and pitch $p$, calculated either by the program MSMS \cite{MSMS}  or using Eq. \eqref{vef} with the $d_{cc}$ distances  reported in table \ref{tab:volumes1}. For comparison, also the value for the LHSC is reported.}
\label{tab:volumes}
\end{table}

\clearpage



\begin{thebibliography}{13}
\bibitem{note1} Note that a finite helix has $C_2$ point symmetry, 
yet we have verified that the helices examined here have nearly uniaxial order 
(with the helix axis as the ordering axis) in the uniaxial nematic phase.
\end{thebibliography}
\end{document}